\documentclass[journal]{IEEEtran}

\usepackage{graphicx}
\usepackage{dcolumn}
\usepackage{bm}
\usepackage[utf8]{inputenc}
\usepackage[T1]{fontenc}
\usepackage{mathptmx}
\usepackage{csquotes}
\usepackage{subcaption}
\usepackage{hyperref}
\usepackage{xcolor}
\hypersetup{linktoc=all}
\usepackage{booktabs}
\usepackage{array}
\usepackage{multirow}
\usepackage{siunitx}
\usepackage[switch]{lineno}

\begin{document}

\title{Exploring Sensitivity of ICF Outputs to Design Parameters in Experiments Using Machine Learning}

\author{\IEEEauthorblockN{Julia~B.~Nakhleh\IEEEauthorrefmark{1}, M.~Giselle~Fern\'andez-Godino\IEEEauthorrefmark{2}, Michael~J.~Grosskopf\IEEEauthorrefmark{1}, Brandon~M.~Wilson\IEEEauthorrefmark{1}, John~Kline\IEEEauthorrefmark{1}, and~Gowri~Srinivasan\IEEEauthorrefmark{1}}\\
\IEEEauthorblockA{\IEEEauthorrefmark{1}Los Alamos National Laboratory, Los Alamos, NM 87544 USA}\\
\IEEEauthorblockA{\IEEEauthorrefmark{2}Lawrence Livermore National Laboratory, Livermore, CA 94550 USA}
\thanks{Corresponding author: Julia B. Nakhleh (email: \href{mailto:jnakhleh@lanl.gov}{jnakhleh@lanl.gov}).}}
\markboth{IEEE Transactions on Plasma Science,~Vol.~X, No.~X,~XXX~XXXX}%
{Nakhleh \MakeLowercase{\textit{et al.}}: Exploring Sensitivity of ICF Outputs to Design Parameters in Experiments Using Machine Learning}

\maketitle

\begin{abstract}
Building a sustainable burn platform in inertial confinement fusion (ICF) requires an understanding of the complex coupling of physical processes and the effects that key experimental design changes have on implosion performance. While simulation codes are used to model ICF implosions, incomplete physics and the need for approximations deteriorate their predictive capability. Identification of relationships between controllable design inputs and measurable outcomes can help guide the future design of experiments and development of simulation codes, which can potentially improve the accuracy of the computational models used to simulate ICF implosions. In this paper, we leverage developments in machine learning (ML) and methods for ML feature importance/sensitivity analysis to identify complex relationships in ways that are difficult to process using expert judgment alone. We present work using random forest (RF) regression for prediction of yield, velocity, and other experimental outcomes given a suite of design parameters, along with an assessment of important relationships and uncertainties in the prediction model. We show that RF models are capable of learning and predicting on ICF experimental data with high accuracy, and we extract feature importance metrics that provide insight into the physical significance of different controllable design inputs for various ICF design configurations. These results can be used to augment expert intuition and simulation results for optimal design of future ICF experiments.
\end{abstract}

\begin{IEEEkeywords}
Inertial confinement fusion, machine learning, random forest.
\end{IEEEkeywords}

\section{Introduction}
\label{sec:introduction}
Inertial confinement fusion (ICF), a technique for generating nuclear fusion reactions by heating and compressing a deuterium-tritium (DT) filled capsule, has been a focus of nuclear fusion research for decades~\cite{betti_inertial-confinement_2016}. Many modern ICF experiments are designed using computer simulations that approximate the real physical processes that occur during capsule implosion. However, when attempting to model applications where the underlying physics is not well understood---as is the case in ICF, where extreme temperatures and pressures exceed $T\geq10^7$~K and $P\geq1$~Mbar, respectively---simulations often perform poorly, and are not always validated by experimental data~\cite{humbird_transfer_2019}. Moreover, ICF experiments are expensive to run, meaning that generating large sets of experimental data to validate simulation results is not always feasible.

Machine learning (ML) offers a novel framework for analyzing data from ICF experiments and simulations. Although the use of ML algorithms in the realm of ICF is relatively new, it has demonstrated some early successes in the field. Deep neural networks (DNNs) and sparse Gaussian process (GP) regression have been used as surrogate models for expensive ICF simulations \cite{humbird_transfer_2019,peterson_zonal_2017,hatfield_using_2019}. Humbird et al.~\cite{humbird_transfer_2019} show that DNN surrogates for low-fidelity ICF simulations can be used with transfer learning. Peterson et al.~\cite{peterson_zonal_2017} and Gopalaswamy et al.~\cite{gopalaswamy_tripled_2019} apply modern statistical and ML approaches to design of ICF experiments, showing novel designs with impressive performance. Shalloo et al.~\cite{shalloo_automation_2020} use similar techniques to optimize parameters of laser wakefield experiments.

Hsu et al.~\cite{hsu_analysis_2020} apply ML regression methods to experimental ICF data (the same dataset described in Section~\ref{sec:dataset}) to analyze relationships between experimental outputs of interest and the predictive capability of several ML regression models with that data. The focus of \cite{hsu_analysis_2020} is on predicting relationships between observables from the experiment to learn scaling laws, rather than assessing the relationship between design parameters and capsule performance. 

In this work, we use an ML feature importance analysis technique called Accumulated Local Effects (ALE)~\cite{apley_visualizing_2020} to quantify the strength of non-linear relationships between controllable design inputs and capsule performance in ICF experiments performed at the National Ignition Facility (NIF). The goal of this importance analysis is to augment the understanding of expert designers and provide insight to improve future designs. To facilitate this analysis, we utilize a random forest (RF) predictor~\cite{breiman_random_2001} to identify systematic relationships between controllable design inputs and experimental outputs from the NIF ICF database. ALE is then applied to the prediction model in order to assess the sensitivity of predicted outputs to the design inputs in order to identify the design features most strongly related to changes in output. We find that when applied to the RF model, ALE is able to identify relationships that are consistent with key ICF design changes over time.

In performing this analysis, it is useful both to analyze the dataset as a whole in order to identify changes in design parameters that are strongly correlated with changes in yield over time, which we do in section~\ref{sec:results}, and to analyze the high and low gas fill density shots individually to better control for the major changes in design configuration between these groups, as we do in section~\ref{sec:individual_analysis_group_I_II}. In theory, it would be useful to analyze each of the four major shot campaigns---low foot (LF), high foot (HF), high density carbon (HDC), and big foot (HDC-BF)---individually to further control for design shifts between each campaign; however, the dataset as it currently exists is too small to meaningfully perform this analysis, as the number of data points for each individual campaign is very limited. Nonetheless, the tools and analysis presented here are novel, and lay the groundwork for new directions of future ICF research.

Section \ref{sec:ml_background} and \ref{sec:dataset} introduce the ML methods and data, respectively, used in this work. Section \ref{sec:results} looks at prediction on the full set of outputs and assesses the importance of design features for prediction across output metrics. Section \ref{sec:individual_analysis_group_I_II} presents individual analyses of low- and high-density hohlraum gas fill shots. We finish by summarizing and discussing future work in Sections \ref{sec:discussion_future_work} and \ref{sec:conclusion}.

\section{Machine Learning Background}
\label{sec:ml_background}

The goal of this work is to analyze the input-output relationships encoded in a prediction model and to explore sensitivity between design parameters and ICF capsule performance. For this purpose, we implement the ALE in Python. As an interpretability metric, ALE is used to communicate \textit{why} an ML model makes the decisions that it does, and, consequently, the extent to which humans can predict the model's results~\cite{miller_explanation_2019, kim_examples_2016}. Interpretability is essential to the safety of many systems (such as driverless cars) and, for scientific applications, necessary in order to extract meaningful scientific knowledge from the model's behavior~\cite{doshi-velez_towards_2017}. For ICF analysis, feature importance rankings are a crucial component of model interpretability because they reflect input-output relationships that can augment understanding of ICF physics.

ALE is a model-agnostic measure that describes the extent to which each feature influences the model's predictions. ALE estimates feature importance by analyzing how much the model's predictions change over a small range of each feature, then averaging and accumulating these differences over the prediction space. Specifically, the ALE main effect of feature $X_i$ is:
\begin{equation}
    \text{ALE}(x_i) = \int_{\text{min}(X_i)}^{x_i} E_{X_{-i}}\left[ \frac{\partial f(\mathbf X)}{\partial X_i} \mid X_i = z \right] dz,
\end{equation}
where $f(\mathbf X)$ is the trained ML predictor, $E_{X_{-i}}\left[ \frac{\partial f(\mathbf X)}{\partial X_i} \mid X_i = z \right]$ is the expected value of the partial derivative of $f(\mathbf X)$ in the $X_i$ direction when $X_i = z$ averaging over the distribution of features other than $i$. The value of the ALE main effect at $x_i$ is found by integrating (accumulating) those expected derivatives from the minimum value of the feature, $\text{min}(X_i)$, to $x_i$. In practice, the expected value of the partial derivative is calculated as an average over the sample data using finite differencing across discrete bins. The variance of this main effect function is then used to provide a scalar metric for the importance of the feature. These variances across features are then normalized to provide a relative importance that parallels variance-based sensitivity analysis like Sobol indices~\cite{saltelli_global_2008}. For further details on the ALE algorithm, we refer the reader to \cite{apley_visualizing_2020}.

To encode the relationships between design parameters and performance in the observed data, we use an RF prediction model. RF regression is an ensemble ML method that employs multiple decision trees to produce highly accurate predictions on medium-to-large datasets. Decision trees are popular due to their efficiency and adaptability, but perform poorly on unseen data~\cite{de_ville_decision_2013, hastie_elements_2009}. RFs reduce over-fitting by averaging over multiple decision trees: each tree is fit to a random sample of the full training data, and for each split of the tree, a random subset of the full features is considered~\cite{breiman_random_2001}. RFs exhibit low generalization error on large datasets, and perform better than individual decision trees on both seen and unseen data~\cite{breiman_random_2001, tin_kam_ho_random_1995}. 

We select RF regression as the predictor based on predictive performance (measured as $R^2$ goodness-of-fit score) on a hold-out test set as shown in Table \ref{tab:model_results_train_test_all_vars}. This result is consistent with the expectation that high performance with DNNs requires more data than this set provides. RFs are also less sensitive to hyperparameter tuning than more complex models such as DNNs. Moreover, RFs are often favored in applications where the underlying intention is to decipher correlations among features in the data. More sophisticated deep ML methods tend to obscure the relationships between features, which makes explainability a topic of great interest among ML practitioners. 

Because ALE is model-agnostic, it is worth noting that it can be used to perform sensitivity analysis on any predictive surrogate model that exhibits good performance in a particular experimental context. In this work, we choose to apply it to an RF model because RF regression exhibits the best performance of all regression methods tested on the given data. However, ALE may prove to be a useful analytical tool for other types of ICF surrogate models as well, not just the RF models studied in this work.

Comparison of feature importance rankings generated by ALE and those extracted from Mean Decrease in Impurity (MDI), the default scikit-learn importance metric for RF models, show that both metrics communicate  similar results and differ only in their rankings of low-ranked variables. We focus here on ALE because of its theoretical advantage with correlated features, which are common in ICF data~\cite{apley_visualizing_2020}, and because of the extra information given by consistent estimation of the main effect.

\section{Dataset}
\label{sec:dataset}
We train our regression model on data from 141 experiments conducted at the NIF beginning in 2011. Our work utilizes 21 design parameters simultaneously in order to predict each of four experimental outputs: total yield, velocity, $\rho R$ from $dsr$ (we refer to this parameter as simply $\rho R$), and gated X-ray bang time (referred to here as BT). Total yield represents the actual measured yield for fusion neutrons, corrected for the small portion of neutrons that lose energy due to scattering as they pass through the ice layer. Velocity (in \SI{}{\micro\meter}/ns or km/s) refers to the implosion velocity. $\rho R$ (in g/\SI{}{\centi\meter\squared}) is a measure of fuel compressibility, calculated as the product of average fuel mass density and fuel radius (assuming a spherical shape for the fuel within the capsule). BT (in ns) is the time at which the fusion neutrons were produced in the experiment, measured according to the time at which X-rays come out of the capsule. All of the inputs and outputs are continuous quantities except for hohlraum material, a discrete quantity whose possible values---gold (Au), uranium (U), and a 75-25 uranium-gold alloy U(Au)---are encoded their respective percentages of uranium, \textit{i.e.} $\text{Au} = 0$, $\text{U} = 1$, and $\text{U(Au)} = 0.75$.

\begin{figure*}
    \centering
    \includegraphics[width=0.7\linewidth]{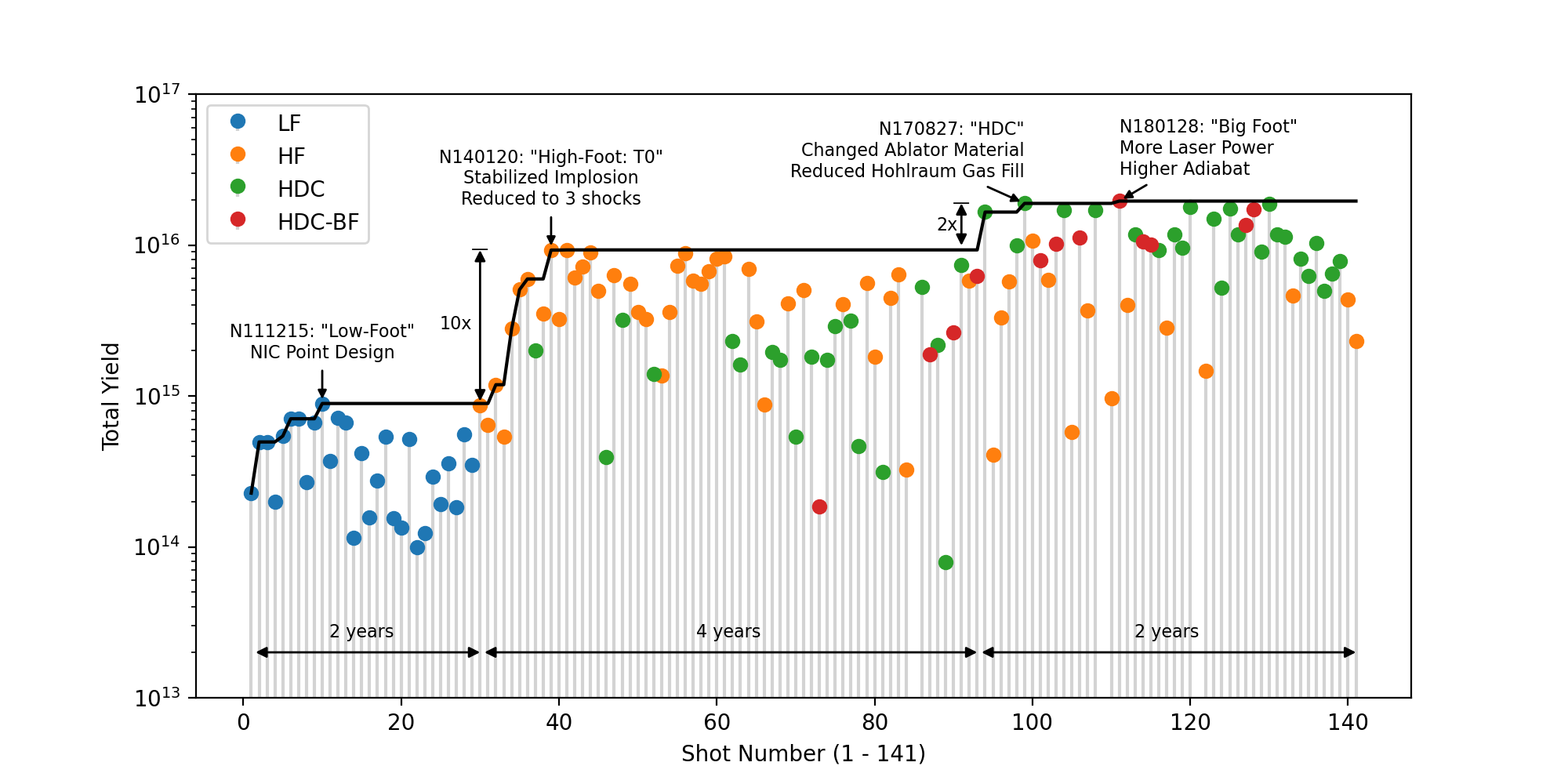}
    \caption{Fusion yield and key design changes for each of the 141 shots carried out at the NIF beginning in 2011.}
    \label{fig:nif_design_changes_over_time}
\end{figure*}

The recorded experiments were performed with a variety of ignition capsule designs and ablator materials. Over this time period, experimental design systematically evolved, resulting in improved performance (see Fig.~\ref{fig:nif_design_changes_over_time}).
In particular, hohlraum design was improved by switching from high density gas fills (group I) to low density gas fills (group II). We define high gas fill density as any value greater than 0.6 mg/\SI{}{\centi\meter\cubed}. Expert opinion and previous work~\cite{hsu_analysis_2020} indicate that, due to the significant physical differences between group I and II shots, separate analysis of each group may improve model prediction and provide insight as to the effects of this design change. In Section~\ref{sec:results}, we analyze RF performance on both groups together, while Section~\ref{sec:individual_analysis_group_I_II} presents individual analyses of each group.

Because the data contains missing values for some features, we pre-process the dataset using iterative imputation from the scikit-learn package in Python~\cite{pedregosa_scikit-learn_2011}, which employs Bayesian ridge regression to estimate (\textit{i.e.} impute) missing values using the remaining observed features. Missingness may arise due to a lack of recording of an actual quantity or because the feature has no natural value for that experiment. The goal of imputation is to provide a value for a missing feature while respecting the underlying relationship between features and performance. In a preliminary analysis, we compared iterative imputation to mean imputation and zero imputation through assessment of the sensitivity of the importance analysis results to the method. Iterative and mean imputation resulted in consistent results, while analysis of results with zero-imputation indicated sensitivity due to several features having imputed values far from the recorded values in the dataset. Hence, we select iterative imputation as the data pre-processing method for the analysis presented in this paper.

The data includes experimental uncertainties for three of the four output quantities studied in this work: total yield, $\rho R$, and BT. The physical origin of the reported uncertainties is not noted in the data; however, we treat each uncertainty measurement as one standard deviation ($\sigma$) to be conservative. The data contains no reported errors for velocity because implosion velocity is not measured directly, but rather inferred via surrogate experiments. Following expert recommendation, we use $\sigma = \pm 15$ \SI{}{\micro\meter}/ns as the uncertainty for all reported values of velocity\footnote{Velocity error is derived from error in the original surrogate convergent ablator experiments and from error in the gated X-ray bang time measurements. The velocity of a DT layered implosion is inferred via a surrogate convergent ablator that uses X-ray radiography to observe capsule radius as a function of time. Using this information along with measured values of X-ray bang time, the velocity of the DT layered implosion can be estimated using a combination of simulations and physics arguments. Using this method, errors in the velocity measurements are principally composed of error in convergent ablator measurement and X-ray bang time measurements, and are typically between 10-15 \SI{}{\micro\meter}/ns.}. Improved quantification of measurement uncertainty would improve the assessment of quality of fit. For all four outputs, we incorporate these uncertainty values into our analysis to provide an indication of whether our model's predictions fall within experimental uncertainty bounds. 

As with any statistical analysis, our results are only as good as the available data. Uncharacterized inputs, such as surface roughness, will not be considered by the ML algorithm. The data also convolves sensitivities to physics mechanisms, such as laser-plasma instabilities and asymmetry and hydrodynamic instability growth with high impact systematic design changes, such as hohlraum design, capsule fill, and laser wavelength tuning. As will be discussed later, ML methods and data analysis methods will identify the most dominant or important features, whether physics-driven or design-driven.

Several of the design parameters recorded in the data are highly correlated, typically the result of deliberate design choices, which are sometimes driven by physical considerations. For example, time-based parameters are correlated with ablator thickness as a design choice, since thicker ablators require a longer push at peak power; as a result, experimenters typically select later times for start final rise and end pulse for shots with thicker ablators. Other correlations are due to multiple variables changing together as part of a broader design change. For instance, hohlraum dimension parameters are highly mutually correlated with one another due to the limited number of hohlraum designs used at NIF over the time period in which the experiments were performed.

Correlated parameters can ``share'' importance in a way that falsely skews importance rankings for any data-driven approach. Following expert recommendation to account for such correlations, the following five input variables were removed from our dataset: start final rise, start peak power, end pulse, Dante 1 diameter, and hohlraum diameter. Ablator thickness and hohlraum length were maintained. A more rigorous assessment of correlated parameters and ML analysis to inform physical relationships is reserved for future work (see Section~\ref{sec:discussion_future_work}).

\section{Results}
\label{sec:results}
We use a random forest model with 100 decision trees and a maximum of four features considered at each tree split. No maximum depth is specified. For total yield, the model trains and predicts on a log scale\footnote{This is done because, unlike other outputs, yield varies across multiple orders of magnitude. Since ALE is a variance-based estimate of feature importance, it overestimates the importance of features that distinguish between orders of magnitude in output. Running the RF with yield on a log scale mitigates this effect.}. The data is not otherwise scaled or normalized.

\subsection{Prediction Quality}
\label{sec:prediction_quality}

\begin{figure*}
\centering
\begin{subfigure}{0.24\textwidth}
  \centering
  \includegraphics[width=\linewidth]{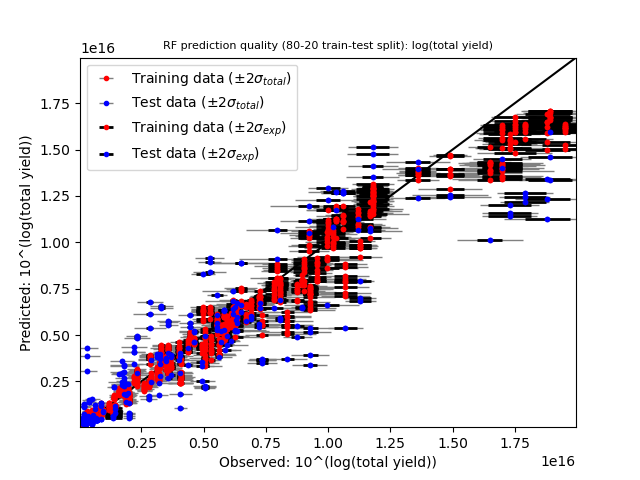}
\end{subfigure}
\begin{subfigure}{0.24\textwidth}
  \centering
  \includegraphics[width=\linewidth]{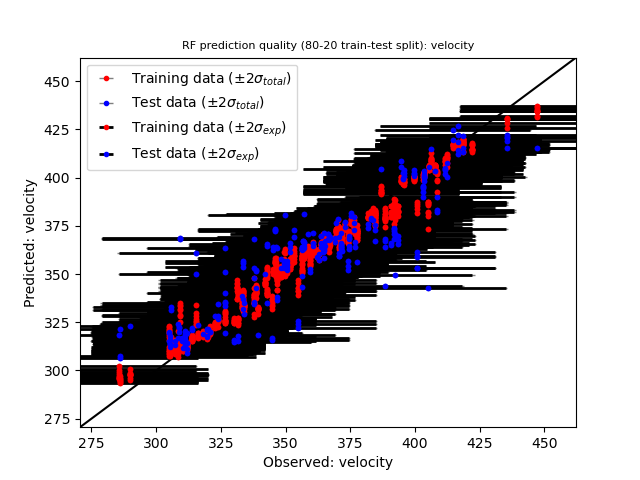}
\end{subfigure}
\begin{subfigure}{0.24\textwidth}
  \centering
  \includegraphics[width=\linewidth]{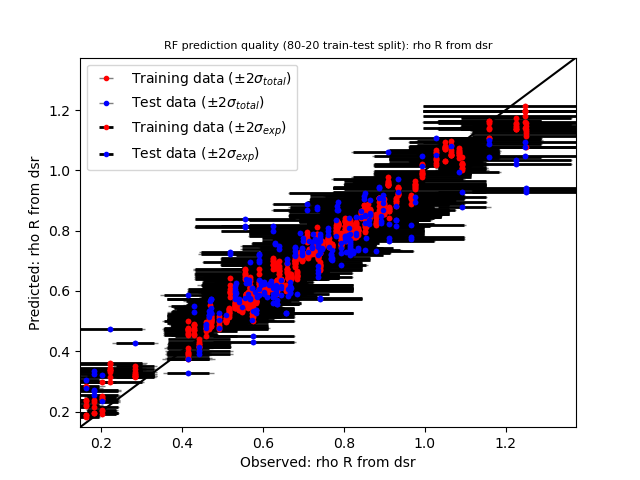}
\end{subfigure}
\begin{subfigure}{0.24\textwidth}
  \centering
  \includegraphics[width=\linewidth]{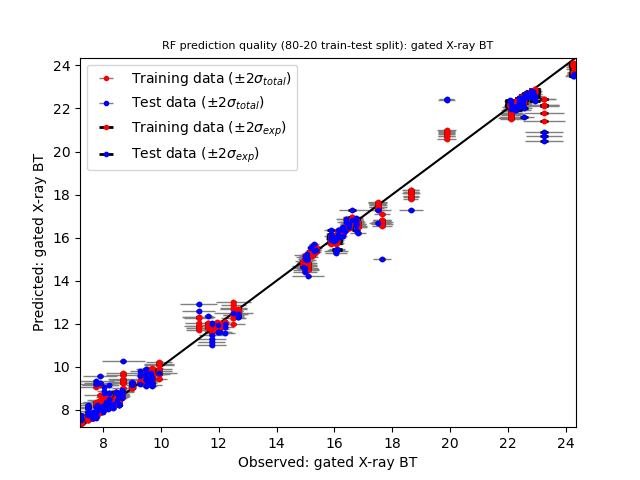}
\end{subfigure}
\caption{Prediction quality for total yield, velocity, $\rho R$, and BT. Each figure displays aggregated results from ten different RF models and random 80-20 train-test splits. Perfect predictions lie along the black $y=x$ line. Black and gray error bars represent $\pm2\sigma_{exp}$ and $\pm2\sigma_{total}$, respectively, where $\sigma_{exp}$ are reported experimental uncertainties and $\sigma_{total}$ are total uncertainties as calculated in Eq.~\ref{sigma_total_eq}. Best viewed in color.}
\label{fig:pred_quality_four_outputs}
\end{figure*}

Fig.~\ref{fig:pred_quality_four_outputs} shows aggregated train-test\footnote{In addition to train and test sets, many ML analyses incorporate an additional validation set, which is used to examine and re-tune the model after initial fitting on the training set but before final evaluation on the test set. Validation sets are used to avoid altering the fitting process after the model has already ``seen'' the test data, as doing so can incorporate experimenter bias into the model. However, due to the small size of our dataset, we do not incorporate a validation set, as doing so would potentially worsen model performance by forcing it to train on an extremely small number of data points. To avoid incorporating experimenter bias into the model, we avoid refitting or changing model parameters based on test data performance.} results for total yield, velocity, $\rho R$, and BT. (Note that for total yield, although the model trains and predicts on a log scale, points are plotted at their original scale.) We show the predictions alongside black and gray error bars representing reported experimental uncertainties and total uncertainties, respectively. Total uncertainty $\sigma_{total}$ is calculated as:
\begin{equation}\label{sigma_total_eq}
    \sigma_{total} = \sqrt{(\sigma_{exp})^2+(\sigma_{model})^2},
\end{equation}
where $\sigma_{exp}$ is the reported experimental uncertainty and $\sigma_{model}$ is the random forest model uncertainty. We estimate $\sigma_{model}$ as $\sigma_{trees}/\sqrt{n}$, where $\sigma_{trees}$ is the standard deviation of the trees' predictions on a given input and $n$ is the number of trees in the forest.

Prediction quality is high across the board, achieving $R^2$ values close to 1 on training data and in the 0.7-0.9 range on test data (see Table~\ref{tab:model_results_train_test_all_vars} for train and test $R^2$ scores on all predicted outputs). Interestingly, the model's predictive quality is particularly high when predicting BT. As an output, BT closely reflects a series of key design changes at the NIF (see Fig.~\ref{fig:nif_design_changes_over_time}). The original LF designs had bang times in the range of ~20 ns, while the newer HF and HDC designs have bang times of approximately 12-14 ns and 8 ns, respectively. For each key design change, yield and implosion velocity have increased while BT has decreased. However, this correlation does not fully explain why the model is able to make such accurate predictions on BT in particular. The concentration of BT values (experimental and predicted) around limited ranges is due to the strong correlation of this parameter with both capsule design and laser pulse. Because only a small number of capsule design and laser pulse configurations were used on the NIF shots, BT values are clustered into a few small groups corresponding with these configurations; however, small variations in capsule thickness and other parameters within the same design configuration create some spread within each cluster. 

The model systematically under-predicts for high experimental values and over-predicts for low ones. This effect may be due to a relative lack of these low and high points in the dataset, as RFs are poor at extrapolating trends for data that they haven't seen during training, and ML models generally perform poorly on test data whose distribution over the parameter space is significantly different from that of training data. However, the bias is visible in the training data as well as the test data, suggesting that the bias at extremes of the dataset is not merely a result of the aforementioned effect, but rather that the given feature space may lack key design features (such as capsule surface quality, mixing between the pusher and the hot and cold fuels, etc.) needed to distinguish medium values of yield, velocity, etc. from very high or low ones.

\begin{figure*}
\centering
\begin{subfigure}{0.24\textwidth}
  \centering
  \includegraphics[width=\linewidth]{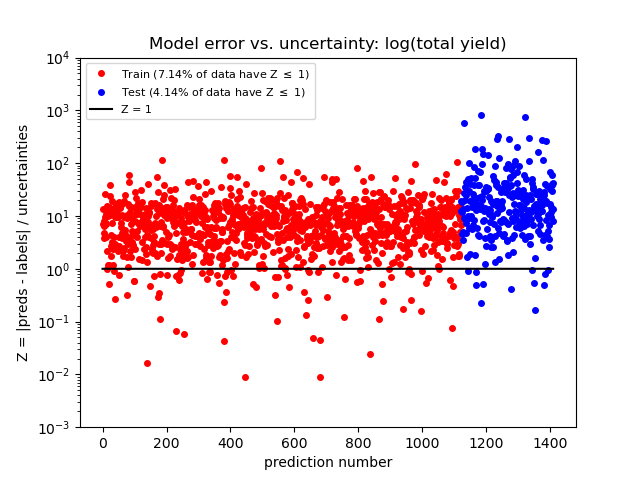}
\end{subfigure}
\begin{subfigure}{0.24\textwidth}
  \centering
  \includegraphics[width=\linewidth]{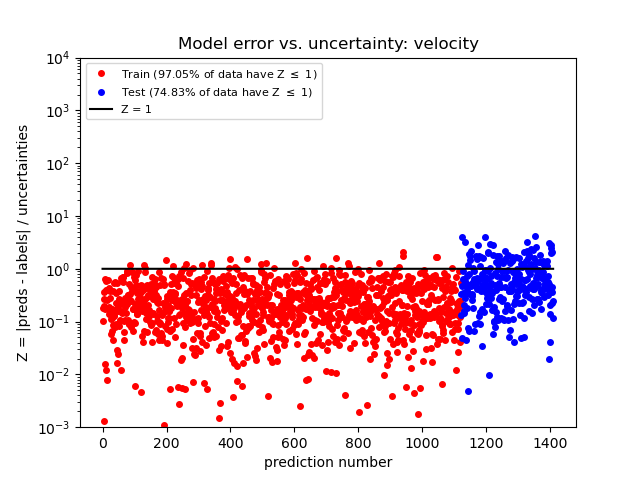}
\end{subfigure}
\begin{subfigure}{0.24\textwidth}
  \centering
  \includegraphics[width=\linewidth]{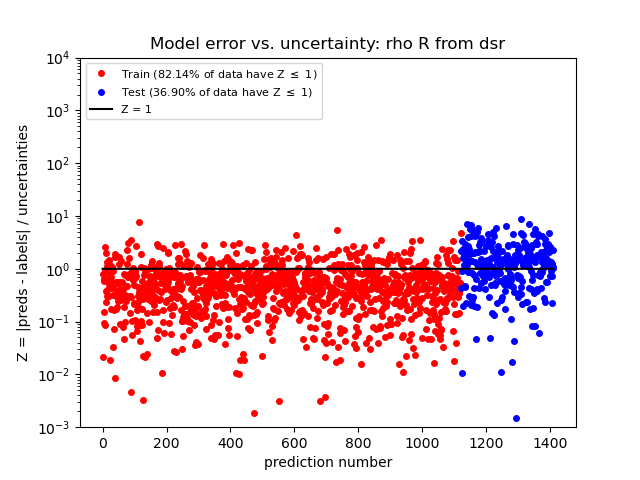}
\end{subfigure}
\begin{subfigure}{0.24\textwidth}
  \centering
  \includegraphics[width=\linewidth]{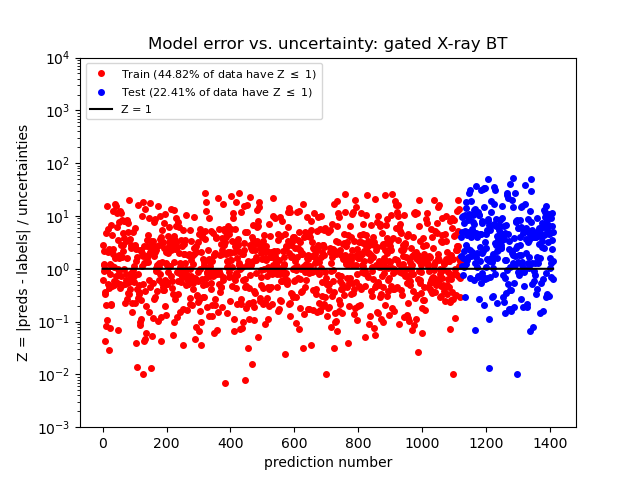}
\end{subfigure}
\caption{Ratio $Z = \frac{|f(X)-Y|}{\sigma_{exp}}$ of model error to experimental uncertainty for total yield, velocity, $\rho R$, and BT. Model predictions that fall within experimental uncertainty bounds fall along or below the black $Z = 1$ line. Best viewed in color.}
\label{fig:error_uncertainty_four_outputs}
\end{figure*}

The ratio of model error to experimental uncertainty---calculated as $Z = \frac{|f(X)-Y|}{\sigma_{exp}}$ where $f(X)$ are model predictions, $Y$ are observed experimental values and $\sigma_{exp}$ are reported experimental uncertainties for $Y$---are shown in Fig.~\ref{fig:error_uncertainty_four_outputs}. The low percentage of points with $Z \leq 1$ for total yield and BT, despite high predictive performance on these values (BT in particular), suggests that the experimental errors reported in the data for total yield and for BT may be too small. For $\rho R$ and velocity, the number of points that fall below the $Z = 1$ line is very high because the reported uncertainties for velocity and $\rho R$ are larger than those reported for total yield. (Velocity and $\rho R$ have average reported experimental percent errors of $4.21\%$ and $6.46\%$, respectively, while the average reported experimental percent errors for total yield and BT are $2.01\%$ and $0.42\%$.)

\begin{table}[ht]
    \centering
    \caption{$R^2$ performance comparison for random forest (RF), Gaussian process (GP), and deep neural network (DNN) models.}
    \label{tab:model_results_train_test_all_vars}
    \begin{tabular}[t]{l>{\centering}p{0.17\linewidth}>{\centering}p{0.15\linewidth}>{\centering}p{0.15\linewidth}>{\centering\arraybackslash}p{0.15\linewidth}}
        \toprule
        Output&&RF&GP&DNN\\
        \midrule
        \multirow{2}{*}{Yield}&Train&\textbf{0.96}&0.87&0.37\\
        &Test&\textbf{0.82}&0.70&0.31\\
        \midrule
        \multirow{2}{*}{Velocity}&Train&\textbf{0.97}&0.87&0.47\\
        &Test&\textbf{0.79}&0.64&0.44\\
        \midrule
        \multirow{2}{*}{$\rho R$}&Train&\textbf{0.97}&0.94&0.52\\
        &Test&\textbf{0.80}&0.77&0.33\\
        \midrule
        \multirow{2}{*}{BT}&Train&\textbf{1.00}&\textbf{1.00}&0.98\\
        &Test&\textbf{0.99}&0.98&0.96\\
        \bottomrule
    \end{tabular}
\end{table}%

\subsection{Importance Results}
\label{sec:importance_results}

\begin{figure}[h]
    \centering
    \includegraphics[width=\linewidth]{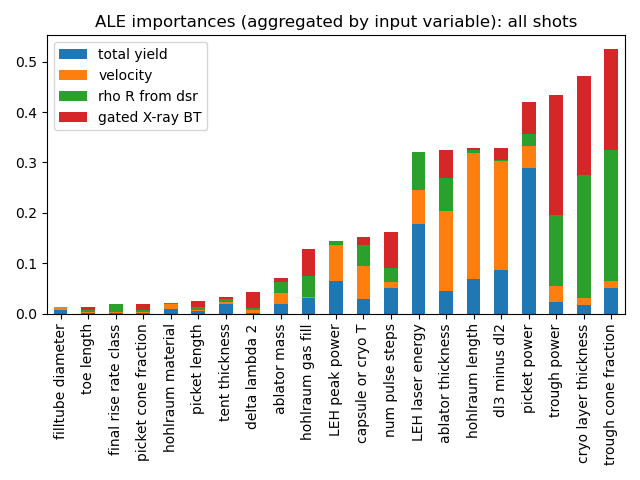}
    \caption{ALE importances, aggregated by input variable in order to visualize which of the 21 inputs are most important to the model's predictions (overall and for each output individually). Best viewed in color.}
    \label{fig:ale_by_input}
\end{figure}

\begin{figure}[h]
    \centering
    \includegraphics[width=\linewidth]{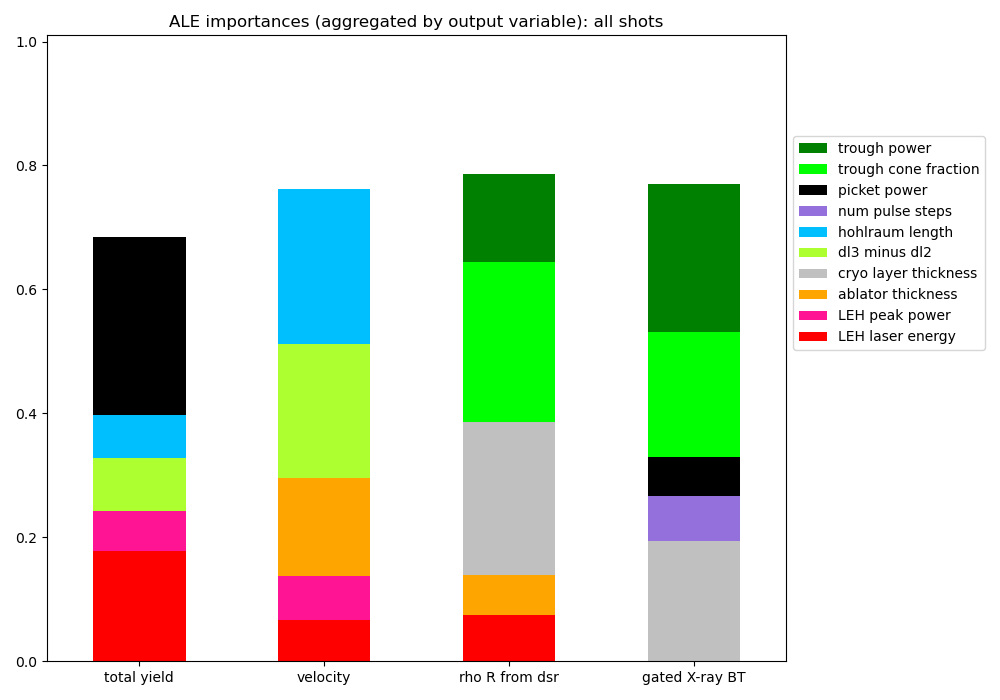}
    \caption{ALE importances, aggregated by output variable in order to visualize which inputs are the strongest predictors of each output. For visual clarity, only the top five inputs contributing to each output are shown. Best viewed in color.}
    \label{fig:ale_by_output}
\end{figure}

We aggregate ALE importance rankings for all outputs by input variable (Fig.~\ref{fig:ale_by_input}) and by output variable (Fig.~\ref{fig:ale_by_output}). Fig.~\ref{fig:ale_by_output} shows that importance rankings between total yield and velocity are highly correlated. High velocity is typically the product of greater kinetic energy in the implosion piston. As the capsule implodes, this energy is deposited into the fuel, creating higher fuel temperature $T$ and greater overall total yield (where yield scales as $T^4$). In Fig.~\ref{fig:ale_by_output}, we see this correlation in the importance rankings for total yield and velocity, both of which show significant effects from $\Delta \lambda_{3} - \Delta \lambda_{2}$, LEH laser energy, and hohlraum length, among other variables. 

Likewise, $\rho R$ and BT show correlated importance rankings: both are strongly influenced by cryo layer thickness, trough power, and trough cone fraction. The correlation between $\rho R$ and BT is likely due to the fact that the original LF ICF designs, which used high gas fill hohlraums and long laser pulses (see Fig.~\ref{fig:nif_design_changes_over_time}), had the largest values of both $\rho R$ and BT. As ICF design shifted toward shorter laser pulse shapes, values of both $\rho R$ and BT decreased. We hypothesize that the high importance of trough cone fraction in predicting both $\rho R$ and BT is due to the fact that trough cone fraction is highly correlated with hohlraum gas fill density (high gas fill shots generally have a longer trough), and gas fill density is an important predictor of implosion performance.

As shown in Fig.~\ref{fig:ale_by_input}, the input variables with the greatest total importance across all outputs are trough cone fraction, cryo layer thickness, trough power, picket power, $\Delta \lambda_{3} - \Delta \lambda_{2}$, hohlraum length, ablator thickness, and LEH laser energy. The high importance of hohlraum length and ablator thickness is likely due to the strong correlation between both variables and the distinct shot campaigns (LF, HF, HDC, and HDC-BF). For each campaign, a limited number of design configurations were tested, each of which used only a few different values of hohlraum length and ablator thickness. Because the largest changes in hohlraum length and ablator thickness occur between different shot campaigns, and these campaigns are a strong predictor of yield, ALE assigns high importance to both hohlraum length and ablator thickness. The issue of correlated variables between design changes presents a challenge for this method of analysis, which we discuss further in Section~\ref{sec:discussion_future_work}. 

Picket power is important for controlling capsule stability during high-speed implosions. Increasing implosion velocity increases energy concentration in the hot spot, thus improving performance and yield~\cite{hurricane_approaching_2019}. However, high-speed implosions typically experience greater instabilities at the ablator surface. Such instabilities, when large enough, reach the hot spot and interfere with neutron reactivity. Increasing the picket power helps reduce such instabilities and prevent them from reaching the hot spot, allowing implosions to be driven stably at higher velocities and thus increasing yield. It is therefore unsurprising that picket power has such high overall importance, particularly in predicting total yield.

The high overall importance of $\Delta \lambda_{3} - \Delta \lambda_{2}$ is likely due to the implosion shape of the first 70 shots (group I). The high density hohlraum gas fill present in these shots causes laser plasma instabilities that make the implosion shape hard to control, causing some of the laser light to scatter back out of the hohlraum and thus reducing implosion yield. The wavelength difference $\Delta \lambda_{3} - \Delta \lambda_{2}$ is used to control the symmetry of high gas fill implosions by controlling the transfer of energy from the outer cone beams to the inner cone beams; however, it can also drive greater backscatter from the inner beams, leading to less overall coupled energy to the target and worse implosion performance overall. The variation in implosion stability, symmetry, and laser backscatter from shot to shot may therefore increase the importance of $\Delta \lambda_{3} - \Delta \lambda_{2}$ when predicting on high gas fill shots, a phenomenon that is not present for the low gas fill shots. Similarly, the high overall importance of trough cone fraction may be the result of correlation between trough cone fraction and hohlraum gas fill, as high gas fill shots generally have a longer trough. 

The important features identified by ALE largely align with the systematic and intentional design changes between the low and high gas fill shots. This finding supports the conclusions of Hsu et al. and highlights the utility of analyzing the low and high gas fill shots independently. We further analyze the discrepancy in feature importance results between both shot groups in Section~\ref{sec:group_I_II_imp_results}.

\section{Individual Analysis of Group I and II Shots}
\label{sec:individual_analysis_group_I_II}

\subsection{Prediction Quality}
\label{sec:group_I_II_pred_quality}

\begin{figure*}
\centering
\begin{subfigure}{0.24\textwidth}
  \centering
  \includegraphics[width=\linewidth]{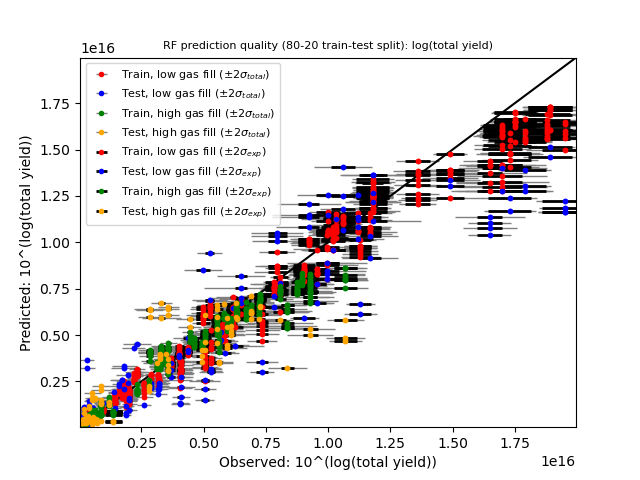}
\end{subfigure}
\begin{subfigure}{0.24\textwidth}
  \centering
  \includegraphics[width=\linewidth]{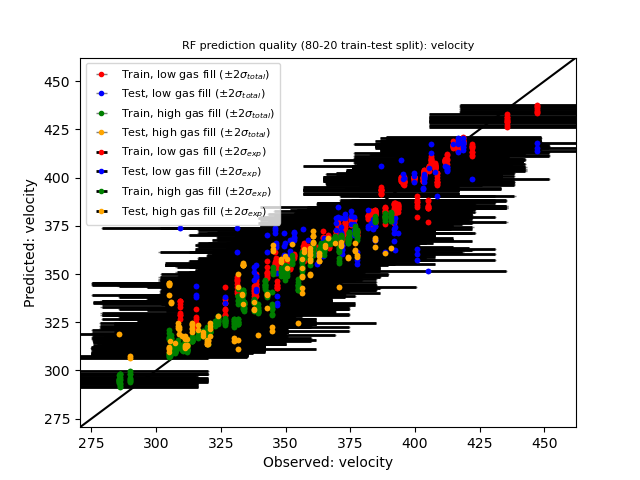}
\end{subfigure}
\begin{subfigure}{0.24\textwidth}
  \centering
  \includegraphics[width=\linewidth]{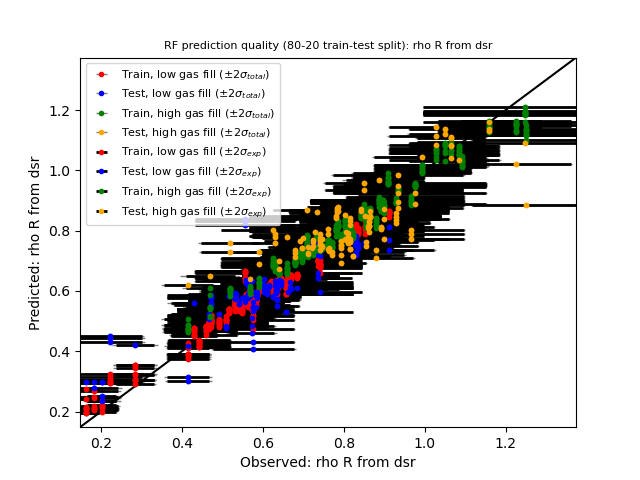}
\end{subfigure}
\begin{subfigure}{0.24\textwidth}
  \centering
  \includegraphics[width=\linewidth]{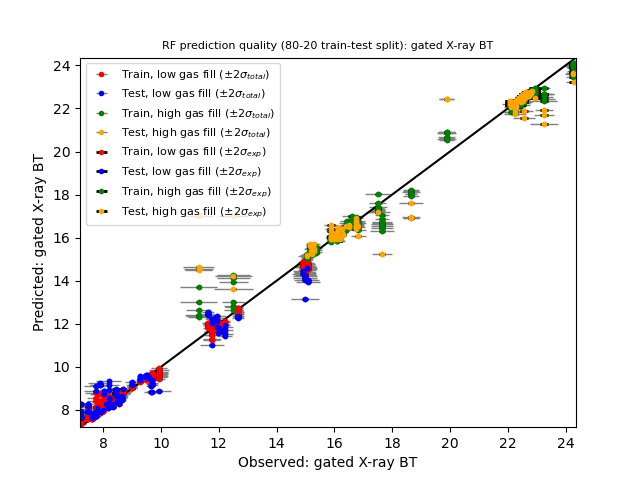}
\end{subfigure}
\caption{Prediction quality for total yield, velocity, $\rho R$, and BT. Each figure displays aggregated results from ten different RF models and random 80-20 train-test splits. Perfect predictions lie along the black $y=x$ line. Black and gray error bars represent $\pm2\sigma_{exp}$ and $\pm2\sigma_{total}$, respectively, where $\sigma_{exp}$ are reported experimental uncertainties and $\sigma_{total}$ are total uncertainties as calculated in Eq.~\ref{sigma_total_eq}. Best viewed in color.}
\label{fig:high_low_pred_quality}
\end{figure*}

Fig.~\ref{fig:high_low_pred_quality} displays aggregated train-test results for high (group I) and low (group II) density shots across all four output variables. Model performance results on are summarized in Table~\ref{tab:model_results_high_low_all_vars}. Again, model performance is generally high, with $R^2$ values close to 1 on training data and in the range of 0.7 to 0.9 on test data. Predictions on total yield and $\rho R$ are slightly higher for the low gas fill shots, while predictions on velocity are slightly higher for the high gas fill shots; however, these differences are extremely slight, and model performance overall is near-equal on both groups. For BT, model performance is worse when predicting on groups I and II individually than when predicting on the dataset as a whole, although prediction quality is still high across the board. As when both groups are analyzed together, the model tends to over-predict low values and under-predict high values for training and test data across all four outputs.

\begin{figure*}
\centering
\begin{subfigure}{0.24\textwidth}
  \centering
  \includegraphics[width=\linewidth]{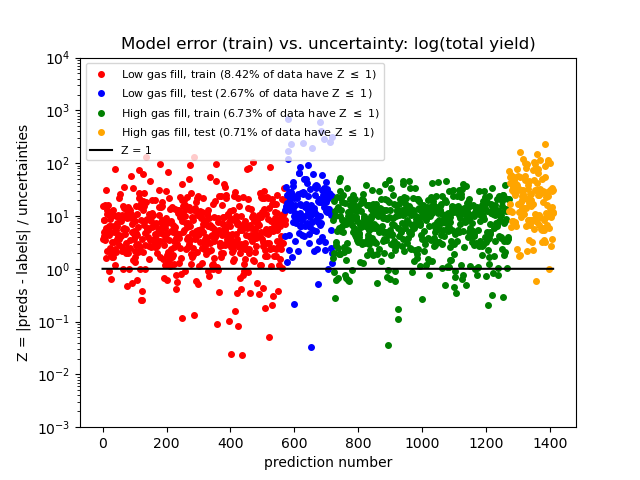}
\end{subfigure}
\begin{subfigure}{0.24\textwidth}
  \centering
  \includegraphics[width=\linewidth]{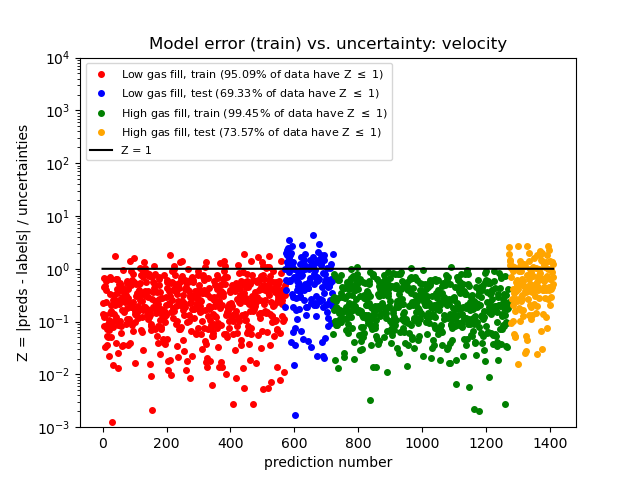}
\end{subfigure}
\begin{subfigure}{0.24\textwidth}
  \centering
  \includegraphics[width=\linewidth]{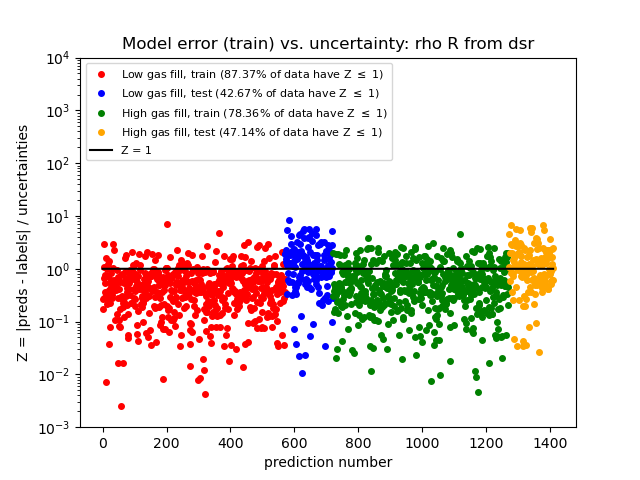}
\end{subfigure}
\begin{subfigure}{0.24\textwidth}
  \centering
  \includegraphics[width=\linewidth]{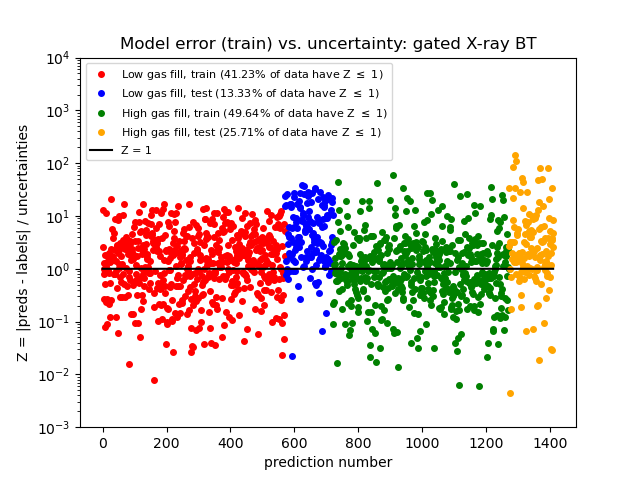}
\end{subfigure}
\caption{Ratio $Z = \frac{|f(X)-Y|}{\sigma_{exp}}$ of model error to experimental uncertainty for total yield, velocity, $\rho R$, and BT. Model predictions that fall within experimental uncertainty bounds fall along or below the black $Z = 1$ line. Best viewed in color.}
\label{fig:high_low_error_uncertainty}
\end{figure*}

Fig.~\ref{fig:high_low_error_uncertainty} displays model error-uncertainty ratio $Z$ for group I and II shots. Again, the majority of data points for total yield have $Z>1$ and the majority of data points for $\rho R$ have $Z \leq 1$. For BT, training data points are split relatively evenly above and below the $Z=1$ line, while a majority of test data points have $Z>1$. For total yield and $\rho R$, the low gas fill data tends to have slightly more points with $Z \leq 1$ than does the high gas fill data, while the opposite is true of velocity. This is consistent with the fact that the model is better on low gas fill data for total yield and $\rho R$ and better on high gas fill data for velocity, although the difference in performance between the two groups is very small. 

\begin{table}[ht]
    \centering
    \caption{$R^2$ model results for high and low gas fill shots.}
    \label{tab:model_results_high_low_all_vars}
    \begin{tabular}[t]{l>{\centering}p{0.14\linewidth}>{\centering}p{0.14\linewidth}>{\centering}p{0.14\linewidth}>{\centering\arraybackslash}p{0.14\linewidth}}
        \toprule
        &Yield&Velocity&$\rho R$&BT\\
        \midrule
        Train (low)&0.96&0.94&0.96&0.99\\
        Test (low)&0.80&0.70&0.70&0.94\\
        Train (high)&0.95&0.96&0.95&0.99\\
        Test (high)&0.74&0.69&0.64&0.92\\
        \bottomrule
    \end{tabular}
\end{table}%

\subsection{Importance Results}
\label{sec:group_I_II_imp_results}

\begin{figure}
    \centering
    \begin{subfigure}{0.45\textwidth}
        \includegraphics[width=\linewidth]{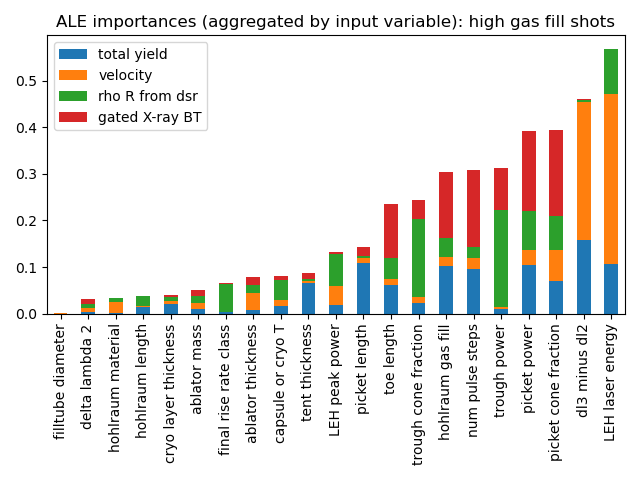}
    \end{subfigure}
    \begin{subfigure}{0.45\textwidth}
        \includegraphics[width=\linewidth]{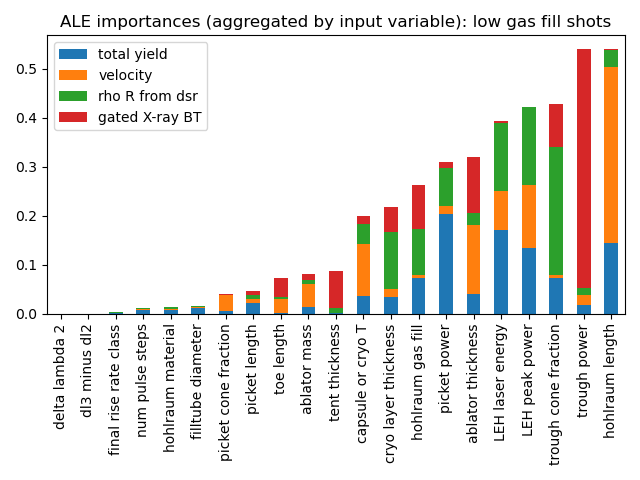}
    \end{subfigure}
    \caption{ALE importances for groups I (top) and II (bottom), aggregated by input variable in order to visualize which of the 21 inputs are most important to the model's predictions (overall and for each output individually). Best viewed in color.}
    \label{fig:high_low_ale_by_input}
\end{figure}

\begin{figure*}
    \centering
    \begin{subfigure}{0.49\textwidth}
        \includegraphics[width=\linewidth]{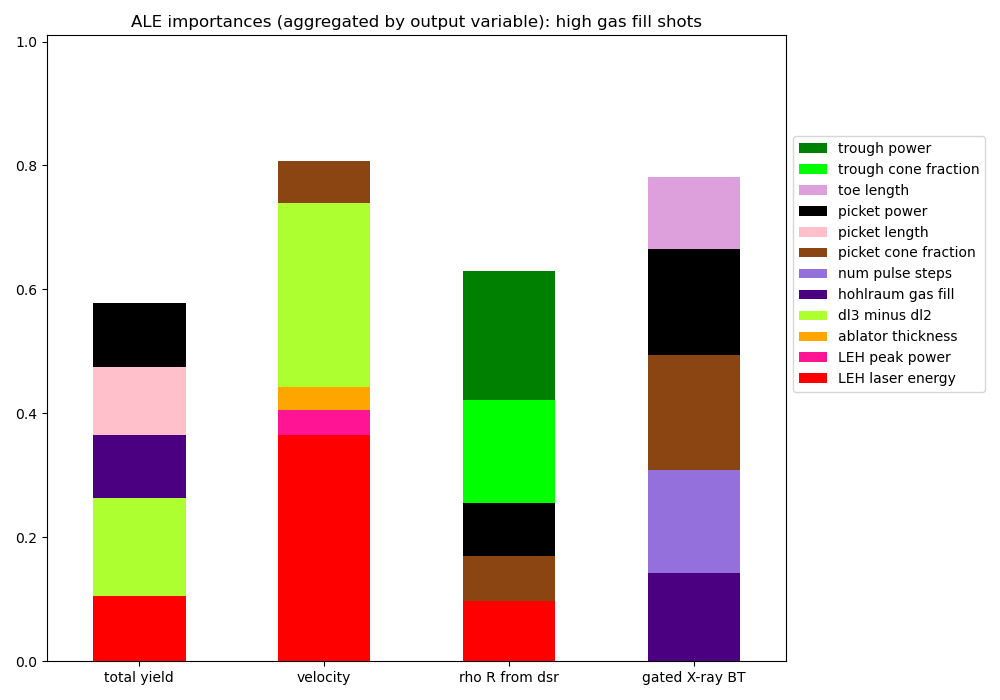}
    \end{subfigure}
    \begin{subfigure}{0.49\textwidth}
        \includegraphics[width=\linewidth]{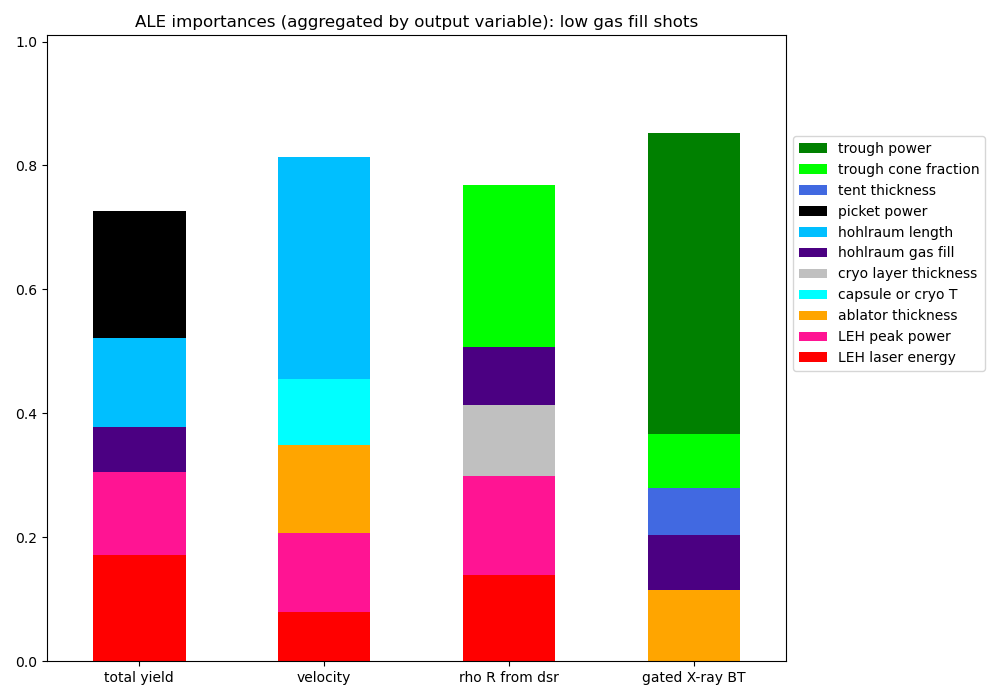}
    \end{subfigure}
    \caption{ALE importances for groups I (left) and II (right), aggregated by output variable in order to visualize which inputs are the strongest predictors of each output. For visual clarity, only the top five inputs contributing to each output are shown. Best viewed in color.}
    \label{fig:high_low_ale_by_output}
\end{figure*}

Fig.~\ref{fig:high_low_ale_by_input} shows importance results for high and low density shots, aggregated by input variable, while Fig.~\ref{fig:high_low_ale_by_output} shows the same results aggregated by output variable. Both figures show significant differences in variable importance rankings between the two shot groups. From Fig.~\ref{fig:high_low_ale_by_input}, we see that, apart from LEH laser energy and trough power, the most important inputs overall for the high gas fill shots are $\Delta \lambda_{3} - \Delta \lambda_{2}$, picket cone fraction, number of pulse steps, picket power, and toe length. In contrast, the low gas fill shots are principally affected by LEH peak power, hohlraum length, trough cone fraction, ablator thickness, and hohlraum gas fill. Notably, the $\Delta \lambda_{3} - \Delta \lambda_{2}$ parameter drop from being the second-most important predictor of high gas fill shots to zero importance for the low gas fill shots. This result is consistent with the fact that for high gas fill shots, the wavelength difference $\Delta \lambda_{3} - \Delta \lambda_{2}$ varies greatly between shots due to laser plasma instabilities caused by the gas fill. With low density hohlraum gas fill, these instabilities are reduced, making $\Delta \lambda_{3} - \Delta \lambda_{2}$ more consistent between shots and thus reducing its predictive importance. 

Trough cone fraction, the most important variable for the dataset as a whole, drops in importance for group I, but remains a significant predictor for group II. Trough cone fraction has a much stronger effect on pulse shape for low gas fill shots, as the trough cone fraction determines how much power can pass by the waist of the capsule before the hohlraum expands to block the lasers from reaching that region (the principal function of the gas fill in early hohlraum designs was to prevent the hohlraum from expanding and blocking the lasers in this manner).

Picket power, the second-most important variable when making predictions on the dataset as a whole, also drops in importance for both groups individually, particularly for group II. This may be because the shift toward low gas fill hohlraums was accompanied by a shift toward more stable HDC-BF implosions, potentially reducing the importance of picket power in setting the fuel adiabat. The largest change in picket power occurred between the LF and HF campaigns, which both used high gas fills; for this reason, the importance of picket power is higher for group I than for group II. Although there was also a shift in picket power between the HDC and BF campaigns (both of which used low gas fills), it was not as significant, resulting in a lower importance ranking for picket power among the low gas fill shots.

We note in Figure \ref{fig:high_low_ale_by_input} that the LEH peak power is notably more important for the group I experiments as compared to group II. Potentially related, the LEH laser energy is of notably higher importance in group I than group II. One possible explanation for this is that the LEH peak power and LEH laser energy are strongly correlated experimentally. As such, the RF predictor can utilize them similarly to capture the same relationship with the yield. Extending the approach in this paper to account for these design-related correlations is the focus of the accompanying paper by Fern\'andez-Godino et al.~\cite{fernandez-godino_identifying_2020}.

From Fig.~\ref{fig:high_low_ale_by_output}, we see that for both shot groups, the importance rankings for total yield and velocity are still correlated to some extent, particularly for the low density shots. For the high gas fill shots, importance rankings between yield and velocity are similar except for the fact that the yield is affected by a greater number of inputs, while velocity is dominated by LEH laser energy and $\Delta \lambda_{3} - \Delta \lambda_{2}$. The strong effect of $\Delta \lambda_{3} - \Delta \lambda_{2}$ on yield outputs and velocity disappears for the low gas fill shots. For low gas fill shots, yield and velocity are mainly affected by LEH peak power, LEH laser energy, hohlraum length, and trough cone fraction. Total yield also shows a lesser, but still significant, effect from picket power, while velocity is strongly affected by ablator thickness and capsule or cryo layer thickness.

Although the importance rankings for $\rho R$ and BT are highly correlated when all shots are analyzed together, this correlation disappears when the two groups are analyzed separately. For the low density shots in particular, the importance rankings for $\rho R$ are more heavily correlated with total yield and velocity than they are with BT. As discussed in Section~\ref{sec:importance_results}, the correlation of $\rho R$ and BT when the dataset is analyzed as a whole is likely due to the correlation of both variables with hohlraum gas fill density. When the data is pre-split by gas fill density, the model may no longer be able to detect this relationship in the data. 

\section{Discussion and Future Work}
\label{sec:discussion_future_work}

In this work, we have demonstrated the use of ML feature importance analysis with ALE to assess strong systematic relationships between design features and ICF capsule performance. At the outset, it was an open question whether the dominant systematic relationships used by the ML predictor model would be driven by physical mechanisms or through other non-causal relationships such as correlation in the data driven by design considerations. While the results show that the sensitivity of capsule performance due to physics mechanisms is strongly confounded with the impact of design changes, the results still communicate important information about systematic structure, or the lack thereof, in the data. Rather than taking the ML importance results thoughtlessly, the results should be used to augment the search for physical understanding by asking the questions: What physical reasons would drive the important systematic relationships to be the way they are, and why is there no systematic relationship between performance features deemed ``unimportant'' by the prediction model?

Expanding the capability to recover the dominant physics mechanisms from data requires addressing the confounding through modeling of correlations among design inputs, new experiments with deliberately randomized statistical design, or reanalysis of current data including assumptions of the causal structure present. For a detailed examination of modeling the correlations between design variables using sparse matrix decomposition, we direct the reader to a companion work by Fern\'andez-Godino et al.~\cite{fernandez-godino_identifying_2020}. Additionally, the strength of many physical relationships can be illustrated through use of simulations or physical experiments specifically targeting those aspects of the system. The potential for inclusion of data from simulation with data from both ICF and wider high-energy-density physics experiments may provide another avenue to adjust for this confounding. Other directions for future work include using ML to analyze simulation-experiment discrepancy and to predict optimal ICF simulation configurations.

One challenge that this work bears out is the fact that the data does not indicate strong relationships between many of the variables, such as those related to the fill tube, tents, and implosion shape, and experimental outputs of interest. This result may indicate that some physical relationships between inputs and outputs are not as strong as is currently thought; for example, an analysis of the laser indirect drive approach at NIF~\cite{town_laser_2020} shows that the changes made to many of the design inputs did not produce changes in yield that are statistically significant, \textit{i.e.} beyond $2\sigma$. This finding indicates that there is missing physics that is not being captured by existing 2D and 3D models; even the best 3D models, for example, do not provide self-consistent estimates of outputs such as yield and ion temperature, which should scale together~\cite{town_laser_2020}. 

However, it is also possible that these physical relationships do exist in the data but are overshadowed in magnitude by the strong effects of design correlations. Because of the limited amount of data available and the large number of design parameters varied simultaneously between experiments, the strongest relationships will be most easily detected by the ALE metric; weaker relationships may exist, but if these are obscured in the data by factors such as the small size of the dataset or variation in output due to extraneous factors, our model will not detect them. Expanding this dataset with data from future experiments, or existing tuning experiments not analyzed here, may help illuminate the more subtle input-output relationships involved in ICF implosions.

\section{Conclusion}
\label{sec:conclusion}
Although its use in the field is relatively new, ML provides numerous promising tools for analyzing ICF data. In this work, we show that RFs are able to learn and predict on data from ICF experiments with high accuracy, achieving $R^2$ scores of 0.9+ on training data and 0.7+ on unseen test data. We also show that the use of ML interpretability metrics can communicate the systematic structure learned by the RF in order to augment the understanding of important relationships in ICF design.  

We build separate RF predictors for high and low gas fill density shots, and show that predictive performance does not differ significantly on either group individually from the performance on the dataset analyzed as a whole. The model's predictions show some bias toward the mean across all outputs and shot groups studied, suggesting that there may be factors missing from the input feature space that affect experimental results. 

Many of the feature importance results detected by our model are consistent with known physics and are reflective of key design changes that took place between experiments. The model's ability to detect shifts in hohlraum gas fill density, capsule design, and other significant changes that took place between shot campaigns indicates its ability to accurately identify which design inputs exert the greatest influence on experimental outputs, providing importance results that are consistent with the physics of ICF implosions. Such ML-based importance results may provide greater insight into input-output relationships as well as the effects of key ICF experimental design changes on outputs of interests, potentially informing the design of future ICF experiments and simulations.

\section{Acknowledgements}
\label{sec:acknowledgements}
This work was supported by the U.S. Department of Energy through the Los Alamos National Laboratory. Los Alamos National Laboratory is operated by Triad National Security, LLC, for the National Nuclear Security Administration of U.S. Department of Energy (Contract No. 89233218CNA000001). We would like to thank Dr. Otto Landen for the use of his NIF experimental database as the source for this work. Approved for public release under LA-UR-20-27991.

\bibliographystyle{IEEEtran}
\bibliography{IEEEabrv,ICF_ML_bibliography}

\end{document}